\documentclass[12pt]{article}
\def\be{\begin{equation}}
\def\ee{\end{equation}}
\def\ber{\begin{eqnarray}}
\def\eer{\end{eqnarray}}
\usepackage{amsfonts,amssymb}
\begin{document}
%\large
\vspace*{1cm}
\begin{center}
{\Large \bf Kaluza-Klein Picture of the World\\[1ex]
and Global Solution of the Spectral Problem}\\

\vspace{4mm}

{\large A.A. Arkhipov\\
{\it State Research Center ``Institute for High Energy Physics" \\
 142280 Protvino, Moscow Region, Russia}}\\
\end{center}

\vspace{4mm}
\begin{abstract}
{We suggest that Kaluza-Klein idea may provide the global solution
of the spectral problem in hadronic spectroscopy.}
\end{abstract}

\section{Introduction}
The strong interactions are characterized by multi-particle
production. The dynamics of the multi-particle systems with a
necessity contains the so called many-body forces. Many-body forces
are fundamental forces which take place in the multi-particle systems
where the number of particles is greater than two, and they are
responsible for the dynamics of the production processes. For
example, the three-body forces are responsible for the dynamics of
one-particle inclusive reactions; see Ref. \cite{1} and references
therein. A description of the many-body forces requires the use of
multidimensional spaces. Therefore, seems it would be naturally to
formulate the strong interactions theory in a multidimensional space
from the beginning.

The idea to use the multidimensional spaces in fundamental physics is
not new: famous works of Kaluza and Klein were the first ones where
this idea has been elaborated. The original idea of Kaluza and Klein
is based on the hypothesis that the input space-time is a
$(4+d)$-dimensional space ${\cal M}_{(4+d)}$ which can be represented
as a tensor product of the visible four-dimensional world $M_4$ with
a compact internal $d$-dimensional space ${\cal K}_d$
\begin{equation}
{\cal M}_{(4+d)} = M_4 \times {\cal K}_d.\label{1}
\end{equation}
The compact internal space ${\cal K}_d$ is space-like one i.e. it has
only spatial dimensions which may be considered as extra spatial
dimensions of $M_4$. An especial example of ${\cal M}_{(4+d)}$ is a
space with the factorizable metric. In according with the tensor
product structure of the space ${\cal M}_{(4+d)}$ the metric may be
chosen in a factorizable form. This means that if $z^M = \{ x^{\mu},
y^{m}\}$, ($M=0,1,\ldots,3+d,\, \mu = 0,1,2,3,\, m=1,2, \ldots, d$),
are local coordinates on ${\cal M}_{(4+d)}$ then the factorizable
metric looks like
\[
ds^{2} = {\cal G}_{MN}(z) dz^M dz^N = g_{\mu \nu}(x) dx^{\mu}
dx^{\nu} + \gamma_{mn}(x,y) dy^{m} dy^{n},
\]
where $g_{\mu \nu}(x)$ is the metric on $M_4$.

In the year 1921, Kaluza proposed a unification of the theory of
gravity and the Maxwell theory of electromagnetism in four dimensions
starting from the theory of gravity in five dimensions. He assumed
that the five-dimensional space ${\cal M}_5$ had to be a product of a
four-dimensional space-time $M_4$ and a circle ${\cal S}_1$: ${\cal
M}_5 = M_4 \times{\cal S}_1$. It was shown that the zero mode sector
of the Kaluza model is equivalent to the four-dimensional theory
which describes the Hilbert-Einstein gravity with a four-dimensional
general coordinate transformations and the Maxwell theory of
electromagnetism with a gauge transformations.

Recently some models with extra dimensions have been proposed to
attack the electroweak quantum instability of the Standard Model
known as hierarchy problem between the electroweak and gravity
scales. However, it is obviously that the basic idea of the
Kaluza-Klein scenario may be applied to any model in Quantum Field
Theory. As example, let us consider the simplest case of
(4+d)-dimensional model of scalar field with the action
\begin{equation}
S = \int d^{4+d}z \sqrt{-{\cal G}} \left[ \frac{1}{2} \left(
\partial_{M} \Phi \right)^2 - \frac{m^{2}}{2} \Phi^2 +
\frac{G_{(4+d)}}{4!} \Phi^4 \right], \label{S}
\end{equation}
where ${\cal G}=\det|{\cal G}_{MN}|$, ${\cal G}_{MN}$ is the metric
on ${\cal M}_{(4+d)} = M_4 \times{\cal K}_d$, $M_4$ is
pseudo-Euclidean Minkowski space-time, ${\cal K}_d$ is a compact
internal $d$-dimensional space with the characteristic size $R$. Let
$\Delta_{{\cal K}_d}$ be the Laplace operator on the internal space
${\cal K}_d$, and $Y_{n}(y)$ are ortho-normalized eigenfunctions of
the Laplace operator
\begin{equation}
\Delta_{K_{d}} Y_{n}(y) = -\frac{\lambda_{n}}{R^{2}} Y_{n}(y),
\label{Yn}
\end{equation}
and $n$ is a (multi)index labeling the eigenvalue $\lambda_{n}$ of
the eigenfunction $Y_{n}(y)$. A $d$-dimen\-sional torus ${\cal
T}^{d}$ with equal radii $R$ is an especially simple example of the
compact internal space of extra dimensions ${\cal K}_d$. The
eigenfunctions and eigenvalues in this special case look like
\begin{equation}
Y_n(y) = \frac{1}{\sqrt{V_d}} \exp \left(i \sum_{m=1}^{d}
n_{m}y^{m}/R \right), \label{T}
\end{equation}
\[
\lambda_n = |n|^2,\quad |n|^2= n_1^2 + n_2^2 + \ldots n_d^2, \quad
n=(n_1,n_2, \ldots, n_d),\quad -\infty \leq n_m \leq \infty,
\]
where $n_m$ are integer numbers, $V_d = (2\pi R)^d$ is the volume of
the torus.

To reduce the multidimensional theory to the effective
four-dimensional one we wright a harmonic expansion for the
multidimensional field $\Phi(z)$
\begin{equation}
\Phi(z) = \Phi(x,y) = \sum_{n} \phi^{(n)}(x) Y_{n}(y). \label{H}
\end{equation}
The coefficients $\phi^{(n)}(x)$ of the harmonic expansion (\ref{H})
are called Kaluza-Klein (KK) excitations or KK modes, and they
usually include the zero-mode $\phi^{(0)}(x)$, corresponding to $n=0$
and the eigenvalue $\lambda_{0} = 0$. Substitution of the KK mode
expansion into action (\ref{S}) and integration over the internal
space $K_{d}$ gives
\begin{equation}
S = \int d^{4}x \sqrt{-g} \left\{ \frac{1}{2} \left(
\partial_{\mu} \phi^{(0)} \right)^{2} - \frac{m^{2}}{2}
(\phi^{(0)})^{2} \right. + \frac{g}{4!} (\phi^{(0)})^{4} +
\end{equation}
\[
+\left. \sum_{n \neq 0} \left[\frac{1}{2} \left(\partial_{\mu}
\phi^{(n)} \right) \left(\partial^{\mu} \phi^{(n)} \right)^{*} -\frac
{m_n^2}{2} \phi^{(n)}\phi^{(n)*} \right] + \frac{g}{4!}
(\phi^{(0)})^{2} \sum_{n\neq 0} \phi^{(n)} \phi^{(n)*}\right\} +
\ldots.
\]
For the masses of the KK modes one obtains
\begin{equation}\label{m}
\fbox{$\displaystyle m_{n}^{2} = m^{2} + \frac{\lambda_{n}}{R^2}$}\,,
\end{equation}
and the coupling constant $g$ of the four-dimensional theory is
related to the coupling constant $G_{(4+d)}$ of the initial
multidimensional theory by the equation
\begin{equation}
\fbox{$\displaystyle  g = \frac{G_{(4+d)}}{V_d}$}\,,\label{g}
\end{equation}
where $V_d$ is the volume of the compact internal space of extra
dimensions ${\cal K}_d$. The fundamental coupling constant
$G_{(4+d)}$ has dimension $[\mbox{mass}]^{-d}$. So, the
four-dimensional coupling constant $g$ is dimensionless one as it
should be. Eqs.~(\ref{m},\ref{g}) represent the basic relations of
Kaluza-Klein scenario.  Similar relations take place for other types
of multidimensional quantum field theoretical models. From
four-dimensional point of view we can interpret each KK mode as a
particle with the mass $m_n$ given by Eq.~(\ref{m}). We see that in
according with Kaluza-Klein scenario any multidimensional field
contains an infinite set of KK modes, i.e. an infinite set of
four-dimensional particles with increasing masses, which is called
the Kaluza-Klein tower. Therefore, an experimental observation of
series KK excitations with a characteristic spectrum of the form
(\ref{m}) would be an evidence of the existence of extra dimensions.
So far the KK partners of the particles of the Standard Model have
not been observed. In the Kaluza-Klein scenario this fact can be
explained by a microscopic small size $R$ of extra dimensions
($R<10^{-17}\,cm$); in that case the KK excitations may be produced
only at super-high energies of the scale $E\sim 1/R > 1\,TeV$. Below
this scale only homogeneous zero modes with $n=0$ are accessible ones
for an observation in recent high energy experiments. That is why,
there is a hope to search the KK excitations at the future LHC and
other colliders.

We have calculated early \cite{2}
\begin{equation}\label{sc}
\frac{1}{R}=41.481 \mbox{MeV},
\end{equation}
or
\begin{equation}\label{size}
\fbox{$\displaystyle R=24.1\,GeV^{-1}=4.75\,10^{-13}\mbox{cm}$}\,.
\end{equation}
If we relate the strong interaction scale with the pion mass
\begin{equation}\label{G}
G_{(4+d)}\sim\frac{10}{[m_\pi]^d},
\end{equation}
then
\begin{equation}\label{simg}
g\sim\frac{10}{(2\pi m_\pi R)^d},
\end{equation}
end
\[
g(d=1)\sim 0.5.
\]
On the other hand
\begin{equation}\label{geff}
g_{eff}=g_{\pi NN}\exp(-m_{\pi}R)\sim 0.5,\ \ \  (g^2_{\pi
NN}/4\pi=14.6).
\end{equation}
So, $R$ has a clear physical meaning: the size (\ref{size}) just
corresponds to the scale of distances where the strong Yukawa forces
in strength come down to the electromagnetic ones. Moreover,
\begin{equation}\label{SM}
M\sim R^{-1}\left(M_{Pl}/R^{-1}\right)^{2/(d+2)}\mid_{d=6}\, \sim
5\,\mbox{TeV}.
\end{equation}
Mass scale (\ref{SM}) is just the scale accepted in the Standard
Model, and this is an interesting observation as well.

\section{Peculiarities of Kaluza-Klein excitations}

From the formula for the masses of the KK modes
\[
 m_{n} = \sqrt{m^{2} + \frac{n^2}{R^2}}
\]\\
one obtains
\begin{equation}\label{potbox}
m_n = m + \delta m_n,\quad \delta m_n = \frac{n^2}{2mR^2},\quad
n<<mR,
\end{equation}
and this just corresponds to the spectrum of potential box with the
size which is equal to the size of internal compact extra space. In
other case we have
\begin{equation}\label{Coulomb}
m_n = n\omega + \delta m_n,\quad \delta m_n =
\frac{m\alpha^2}{2n},\quad \omega \equiv \frac{1}{R},\quad \alpha^2
\equiv mR,\quad \alpha^2<<n,
\end{equation}
and here we come to the (quasi)oscillator (quasi, because $n$ instead
of $n+1/2$ for one-dimensional case) and (quasi)Coulomb (quasi,
because $1/n$ instead of $1/n^2$ and $\alpha^2=mR$ instead of
$\alpha_c^2=(1/137)^2$) spectra. Clearly, we can neglect the
(quasi)Coulomb contribution in the region $n>>\alpha^2\equiv mR$.

It is a very remarkable fact that KK modes of relativistic origin,
being made with a quantization of finite moving in the space of extra
dimensions, interpolate the non-relativistic spectrum of a potential
box and the oscillator spectrum.

The spectrum of two($a$ and $b$)-particle compound system is defined
in fundamental (input) theory by the formula
\begin{equation}\label{comp}
M_{ab}^n = m_a + m_b + \delta m_{ab}^n(m_a,m_b,G_{4+d}).
\end{equation}
The goal of the fundamental theory is to calculate $\delta
m_{ab}^n(m_a,m_b,G_{4+d})$. We have not the solution of that problem
in strong interaction theory because this is significantly
non-perturbative problem. However, in the framework of Kaluza-Klein
approach we can rewrite the above formula in an equivalent form
\begin{equation}\label{KKcomp}
M_{ab}^n = m_{a,n} + m_{b,n} + \delta m_{ab,n}(m_{a,n},m_{b,n},g),
\end{equation}
where $m_{a,n},m_{b,n}$ are KK modes of particles $a$ and $b$, and we
can calculate  using four-dimensional perturbation theory for
quantity $\delta m_{ab,n}(m_{a,n},m_{b,n},g)$. Moreover, because
$\delta m_{ab,n}(m_{a,n},m_{b,n},g)<<m_{a(b),n}$, we can put with a
high accuracy
\begin{equation}\label{approx}
M_{ab}^n \cong m_{a,n} + m_{b,n},
\end{equation}
and this fact allows one to formulate a global solution of the
spectral problem in hadronic spectroscopy.

\section{On global solution of the spectral problem}

According to Kaluza and Klein  we suggest that the input
(fundamental) space-time ${\cal M}_{(4+d)}$ is represented as
\[
{\cal M}_{(4+d)} = M_4 \times {\cal K}_d.
\]
Let $\lambda_{n}$ are characteristic numbers of the Laplace operator
on ${\cal K}_{d}$ with a characteristic size $R_{\cal K}$
\[
\Delta_{{\mathcal K}_{d}} Y_{n}(y) = -\frac{\lambda_{n}}{R_{\cal
K}^{2}} Y_{n}(y).
\]\\
Let $\lambda_{\cal K}$ be the set of all characteristic numbers of
the Laplace operator
\begin{equation}\label{charset}
\lambda_{\cal K} \equiv \biggl\{ \lambda_n: n\in {\mathbb Z}^d \equiv
\underbrace{{\mathbb Z}\times{\mathbb Z}\times\cdots\times{\mathbb
Z}}_d\, \biggr\}.
\end{equation}
There is one-to-one correspondence
\[
{\cal K} \quad \Longleftrightarrow \quad (R_{\cal K},\lambda_{\cal
K}).
\]
Let us consider a compound hadronic system $h$ which may decay into
some channel
\begin{equation}\label{channel}
h \rightarrow a+b+\cdots +c.
\end{equation}
We introduce the spectral mass function of the given channel by the
formula
\begin{equation}\label{spmass}
M_h^{ab...c}(R_{\cal
K},\lambda_{n_a},\lambda_{n_b},\cdots,\lambda_{n_c})=\sqrt{m_a^2 +
\frac{\lambda_{n_a}}{R_{\cal K}^2}} + \sqrt{m_b^2 +
\frac{\lambda_{n_b}}{R_{\cal K}^2}} + \cdots + \sqrt{m_c^2 +
\frac{\lambda_{n_c}}{R_{\cal K}^2}}.
\end{equation}
Now we build the Kaluza-Klein tower:
\begin{equation}\label{tower}
t_h^{ab...c}({\cal K})=t_h^{ab...c}(R_{\cal K},\lambda_{\cal
K})\equiv \biggl\{M_h^{ab...c}(R_{\cal
K},\lambda_{n_a},\lambda_{n_b},\cdots,\lambda_{n_c}):
\lambda_{n_i}\in\lambda_{\cal K} \biggr\},
\end{equation}
\[
(i=a,b,...,c).
\]
After that we build the Kaluza-Klein town as a union of the
Kaluza-Klein towers corresponding to all possible decay channels of
the hadronic system $h$
\begin{equation}\label{town}
{\cal T}_h({\cal K}) = {\cal T}_h(R_{\cal K},\lambda_{\cal K})\equiv
\bigcup_{\{ab...c\}}t_h^{ab...c}(R_{\cal K},\lambda_{\cal K}).
\end{equation}

We state:
\begin{equation}\label{hmass}
\fbox{$\displaystyle M_h \in {\cal T}_h({\cal K})$}\,.
\end{equation}
Let ${\cal H}$ be the set of all possible physical hadronic states.
We build the hadronic Kaluza-Klein country by the formula
\begin{equation}\label{country}
{\mathbb C}_{\cal H}({\cal K}) \equiv \bigcup_{h\in {\cal H}}{\cal
T}_h({\cal K}).
\end{equation}
The whole spectrum of all possible physical hadronic states we denote
$M_{\cal H}$
\begin{equation}\label{spectr}
M_{\cal H} \equiv \biggl\{M_h: h\in {\cal H}\biggr\}.
\end{equation}

We state:
\begin{equation}\label{globsol}
\fbox{$\displaystyle M_{\cal H} \in {\mathbb C}_{\cal H}({\cal
K})$}\,.
\end{equation}
The formulae (\ref{hmass}) and (\ref{globsol}) provide the global
solution of the spectral problem in hadronic spectroscopy.

Here we have to make some clarifying remarks. First of all, in the
construction of the global solution among all possible decay channels
of the hadronic system $h$ there have to be taken into account only
those channels which contain the fundamental particles and their
different multi-particle compound systems in the final states, as it
should be. An appearance of non-zero KK modes of the fundamental
particles and their compound systems in the final states of the decay
channels is forbidden by the construction. For example, the decay
channel
\begin{equation}\label{channel2}
h \nrightarrow a^*+b+\cdots +c,
\end{equation}
where $a^*$ is a non-zero KK mode of the fundamental particle $a$,
cannot be used in the construction. The decay channel
\begin{equation}\label{channel3}
h \rightarrow A+b+\cdots +c,
\end{equation}
where $A$ is some multi-particle compound system which may decay into
some channel with the fundamental particles $a_i (i=1,2,...k)$ in the
final state
\begin{equation}\label{channel4}
A \rightarrow a_1+a_2+\cdots +a_k,
\end{equation}
is admissible one by the construction. But the decay channel
\begin{equation}\label{channel5}
h \nrightarrow A^*+b+\cdots +c,
\end{equation}
where $A^*$ denote some non-zero KK mode of $A$, is forbidden. In
other words, the underlying physical principle in the construction of
the global solution was the principle of non-observability of
non-zero KK modes of the fundamental particles and their compound
systems. According to that principle non-zero KK modes of the
fundamental particles may manifest themselves only virtually during
an interaction, for example, when they are staying in a compound
system. Non-zero KK modes of the fundamental particles living in a
compound system define the main properties of a compound system such
as the mass and the life time of the system. As mentioned above, an
interaction of KK modes is weak, therefore we can calculate with a
high accuracy the mass of a compound system as a simple sum of the
masses of KK modes. Moreover, weakly interacting KK modes result very
narrow widths of the compound states, and this phenomenon is observed
at the recent experiments. If we say about the broad peaks in the
hadronic spectra we interpret them as an envelope of the narrow peaks
predicted by Kaluza-Klein scenario. Really, the dynamics of the
compound systems decays is physically transparent: Non-zero KK modes
of the constituents make a transition to zero KK modes, and we
observe zero KK modes as the decay products.

We shown in the previous section that non-zero KK modes look like the
states of a particle in confine potentials. Such particle might be
considered as a quasi-particle which cannot be observed without the
destroying a confine potential. A quasi-particle becomes a real
particle by a transition of a non-zero KK mode to a zero KK mode
which is equivalent to the destroying a confine potential, and we
observe a zero KK mode i.e. real fundamental particle as a decay
product. This consideration justifies the underlying physical
principle in the construction of the global solution. In fact, we
present here quite a new look on the Kaluza-Klein picture as a whole.

\section{One Comment}

In papers \cite{2,3,4,5,6,7} we have verified this global solution on
the set of experimental data with two-nucleon system, two-pion
system, three-pion system, strange mesons, charmed and
charmed-strange mesons and found out that the solution accurately
described the experimentally observed hadronic spectra. At that we
have used the simplest form of torus for the internal compact extra
space and considered only diagonal elements in Kaluza-Klein towers.
In fact, we have established the non-trivial physical principle
according to which KK modes of decay products preferably paired up in
compound system when they lived on one and the same storey in
Kaluza-Klein tower. However, there are an exceptional cases. For
example, $\rho$ and $\omega$ mesons appear as non-diagonal elements
of the Kaluza-Klein towers:
\begin{equation}\label{rho}
m_\rho\in M_{n,m}^{\pi^1\pi^2}= \sqrt{m_{\pi^1}^2+\frac{n^2}{R^2}} +
\sqrt{m_{\pi^2}^2+\frac{m^2}{R^2}},
\end{equation}
\[
M_{n,m}^{\pi^+\pi^-}(n_{\pi^+}=12,
m_{\pi^-}=4)=766.97\mbox{MeV},\quad
M_{n,m}^{\pi^+\pi^-}(n_{\pi^+}=13, m_{\pi^-}=4)=773.85\mbox{MeV},
\]
\[
M_{n,m}^{\pi^0\pi^0}(n=13, m=4)=769.78\mbox{MeV},\quad
M_{n,m}^{\pi^+\pi^0}(n_{\pi^+}=13, m_{\pi^0}=4)=770.92\mbox{MeV},
\]
and
\begin{equation}\label{omega}
m_\omega\in M_{n,m,k}^{\pi^+\pi^-\pi^0}=
\sqrt{m_{\pi^+}^2+\frac{n^2}{R^2}} +
\sqrt{m_{\pi^-}^2+\frac{m^2}{R^2}} +
\sqrt{m_{\pi^0}^2+\frac{k^2}{R^2}},
\end{equation}
\vspace{2mm}
\[
M_{n,m,k}^{\pi^+\pi^-\pi^0}(n_{\pi^+}=5,m_{\pi^-}=6,k_{\pi^0}=5)=782.80\mbox{MeV}.
\]
In general, as it follows from the observed hadron spectrum, the
non-diagonal elements of the Kaluza-Klein towers are physically
suppressed.

\section{Conclusion}

We would like to especially emphasize that one simple formula with
one fundamental constant described 120 experimentally observed
hadronic states which distributed as 43 two-nucleon states, 29
two-pion states, 9 three-pion states, 25 strange states and 14
charmed and charmed-strange states. Obviously, this is an impressive
fact. New recent experimental data presented at last Xth
International Conference on Hadron Spectroscopy HADRON '03 (August
31--September 6, 2003, Aschaffenburg, Germany) are in excellent
agreement with the global solution constructed here, and this is a
very impressive fact as well \cite{8}.

The architecture of the hadronic Kaluza-Klein towns is unambiguously
defined by an internal compact extra space with its geometry and
shapes, and we have to learn much more about the geometry and shapes
of a compact internal extra space. However, one very important point
in Kaluza-Klein picture is established now in a reliable way: The
size of the internal compact extra space define the global
characteristics of the hadronic spectra while the masses of the
constituents are the fundamental parameters of the compound systems
which the elements of the global structures being. A knowledge of the
true internal compact extra space is a knowledge of the Everything
that is the God. Our consideration made above shown that we found out
a good approximation to the true internal extra space. In our opinion
the experimentally observed hadronic spectra reveal the existence of
extra dimensions and comfirm the Kaluza-Klein picture of the world,
which allow us to construct the global solution of the spectral
problem in hadronic spectroscopy.


\begin{thebibliography}{**}
\bibitem{1}
A.A.~Arkhipov, hep-ph/0211449 (2002); preprint IHEP 2002-44,
Protvino, 2002, available at
http://dbserv.ihep.su/\~{}pubs/prep2002/ps/2002-44.pdf
\bibitem{2}
A.A.~Arkhipov, hep-ph/0208215 (2002); preprint IHEP 2002-43,
Protvino, 2002, available at
http://dbserv.ihep.su/\~{}pubs/prep2002/ps/2002-43.pdf
\bibitem{3}
A.A.~Arkhipov, hep-ph/0302164 (2003).
\bibitem{4}
A.A.~Arkhipov, hep-ph/0302213 (2003).
\bibitem{5}
A.A.~Arkhipov, hep-ph/0304014 (2003).
\bibitem{6}
A.A.~Arkhipov, hep-ph/0305167 (2003).
\bibitem{7}
A.A.~Arkhipov, hep-ph/0306237 (2003).
\bibitem{8}
A.A.~Arkhipov, hep-ph/0309327 (2003).
\end{thebibliography}
\end{document}